# Extreme adaptive optics imaging with a clear and well-corrected off-axis telescope sub-aperture

E. Serabyn, K. Wallace, M. Troy, B. Mennesson, P. Haguenauer, R. Gappinger
Jet Propulsion Laboratory, California Institute of Technology, Pasadena, CA 91109

and R. Burruss
Palomar Observatory, California Institute of Technology,
PO Box 200, Palomar Mountain, CA 92060

**Abstract**

Rather than using an adaptive optics (AO) system to correct a telescope's entire pupil, it can instead be used to more finely correct a smaller sub-aperture. Indeed, existing AO systems can be used to correct a sub-aperture 1/3 to 1/2 the size of a 5-10 m telescope to extreme adaptive optics (ExAO) levels. We discuss the potential performance of a clear off-axis well-corrected sub-aperture (WCS), and describe our initial imaging results with a 1.5 m diameter WCS on the Palomar Observatory's Hale telescope. These include measured Strehl ratios of 0.92-0.94 in the infrared (2.17 µm), and ≈ 0.12 in the B band, the latter allowing a binary of separation 0.34" to be easily resolved in the blue. Such performance levels enable a variety of novel observational modes, such as infrared ExAO, visible-wavelength AO, and high-contrast coronagraphy. One specific application suggested by the high Strehl ratio stability obtained (1%) is the measurement of planetary transits and eclipses. Also described is a simple "dark-hole" experiment carried out on a binary star, in which a comatic phase term was applied directly to the deformable mirror, in order to shift the diffraction rings to one side of the point spread function.

**Subject headings:** adaptive optics, coronagraphy

Direct detection of faint brown dwarfs and planets close to bright stars requires very high contrast observations at small angular separations. A number of recently-proposed coronagraphs can potentially be used to reduce the bright starlight (e.g. Guyon 2006), but scattering by non-ideal telescope optics and atmospheric seeing fluctuations severely limits the off-axis detection capabilities of all types of coronagraph, with performance ultimately set by the quality of the corrected stellar wavefront (Malbet, Yu and Shao 1995). Indeed, classical "Lyot" coronagraphs employing opaque focal-plane starlight blockers should yield significant contrast gains only for stellar Strehl ratios exceeding ≈ 90% (Sivaramakrishnan et al. 2001). On the other hand, current-generation AO systems on large ground-based telescopes typically provide Strehl ratios of ≈ 50 - 70%, limiting coronagraphic performance.

As we suggest here, a higher degree of wavefront correction can be obtained by using an existing wavefront sensor (WFS) and deformable mirror (DM) to more finely sample and correct a telescope sub-aperture smaller than the full telescope pupil. In this case, even with a lower stellar flux per WFS element, the smaller effective DM spacing can still yield significantly better wavefront correction across the subaperture. Alternatively, a current-generation AO system could be installed on a smaller telescope, but imaging and coronagraphic performance are also degraded by the pupil blockage due to an on-axis secondary mirror and its supports. Since an off-axis sub-aperture can eliminate vignetting, a clear, off-axis, "well-corrected sub-aperture" (WCS) can provide a dual image improvement - a reduction in the scattering due to phase aberrations, and the elimination of scattering by vignetting elements. Note that such a WCS is quite easy to implement – a set of relay optics is simply inserted prior to an AO system (Fig. 1) to magnify and shift a sub-aperture image onto the DM (Haguenauer et al. 2005, 2006; Serabyn et al. 2006).

Although both of the factors mentioned impact image quality, the commonly used measure of image quality, the Strehl ratio, S, reflects only one of these. S is normally defined in the presence of pupil plane phase errors as the ratio of the intensity at the center of the aberrated focal-plane point spread function (PSF) to the intensity at the center of the ideal, unaberrated PSF (which is given by the Fourier transform of the pupil.) For a pupil-plane wavefront with an rms phase error of $\phi$ radians, the familiar relationship

$$S = e^{-\phi^2} \qquad (1)$$

applies (e.g. Schroeder 1987). This approach takes pupil vignetting into account only insofar as the ideal or "reference" PSF depends on the shape of the partially obstructed aperture. On the other hand, even though vignetting elements such as a secondary mirror and secondary supports modify the ideal PSF by transferring light from the core of the PSF into the PSF wings, the Strehl ratio is not affected. Thus, the standard Strehl ratio cannot be used to directly compare the imaging performance of clear and partially vignetted apertures.

It is however quite straightforward to compare these two cases on an equal footing, using the same discriminator as that on which the Strehl ratio is based – the central PSF intensity. Indeed, for an aperture with a small fractional areal obscuration, f, Babinet's theorem (Jackson 1975) implies that the central image plane intensity is fractionally reduced by 2f, half by direct absorption, reflecting a throughput of 1-f, and half by scattering, which increases the scattered light level by f, and decreases the central intensity by another factor of 1-f. Thus, referred to the light transmitted by the partially vignetted aperture, the factor 1-f accounts quantitatively for the "scattering" due to pupil vignetting in the same manner that S describes the scattering by phase errors. In the limit of small blockage and phase aberration, the ratio of the PSF peak for a given vignetted aperture to the PSF peak for an ideal clear and aberration-free pupil of the same outer diameter (where each PSF is normalized to its relevant transmitted flux) is then simply given by the product of two individual scattering terms, or

$$S' = (1-f)\,S. \qquad (2)$$

Thus, for a partially vignetted pupil, S remains unchanged from the clear aperture case, but S' is reduced by 1-f, accurately reflecting the higher light level in the PSF wings[1]. In the unvignetted case, 1-f = 1, while for vignetted pupils this factor typically lies in the range 0.8 – 0.99.

The level of wavefront improvement that can be provided by a WCS is illustrated in Fig. 2, which compares Strehl ratios obtainable for the full aperture and sub-aperture cases for the Palomar Observatory's 200-inch diameter Hale telescope, and for one of the Keck telescopes on Mauna Kea (modeled for simplicity as a 10 m diameter telescope), as a function of $r_o$, the atmospheric seeing cell size. For typical $r_o$'s of 30 - 50 cm in the K band (2.2 μm), a WCS at Palomar can reduce the rms wavefront error from the currently typical 200 - 250 nm to just under 100 nm, thus increasing S from ≈ 40-70% to 88-96%. The Strehl ratios for the larger telescope would not be quite as high because of the larger effective actuator spacing (≈ 20 cm vs. 10 cm) available with current DMs, but future DMs will likely reduce this difference.

Of course at a given wavelength, λ, a WCS provides worse angular resolution than the full telescope pupil. This resolution loss is potentially offset by the ability to make coronagraphic observations closer to the star in λ/D units, where D is the sub-aperture diameter, because of the improved wavefront in the sub-aperture (Sivaramakrishnan et al. 2001), especially if use is made of a coronagraph with small intrinsic inner working angles. Indeed, the four-quadrant phase mask (FQPM) coronagraph actually requires a clear circular pupil for high rejection (Rouan et al. 2000, Riaud et al. 2001, Lloyd et al. 2003), and so is a natural match to an unvignetted off-axis optical system. The coronagraphic potential of a WCS will be explored in a subsequent paper; in this paper we focus first on the improved imaging performance achievable with a WCS.

A WCS can also provide improved performance at shorter wavelengths, thus providing a means of maintaining angular resolution. Indeed, wavefront errors < 100 nm rms provide good Strehl ratios well into the optical (Fig. 3), thus enabling visible-wavelength AO imaging of a quality comparable to that currently available only in the infrared. The ideal wavelength for adaptively corrected imaging is of course subjective (and spectrum dependent), but we may define as an objective criterion the "image concentration," $C$, defined here as the Strehl ratio divided by the approximate beam area,

$$C = \frac{S}{q^2}. \tag{3}$$

where $q$ is the beam half-power width. The maximum image concentration occurs at the zero of the derivative of $C$ with respect to $\lambda$. With $q \approx \lambda/D$, and the rms phase entering into $S$ given by $\phi = 2\pi\sigma/\lambda$, where $\sigma$ is the pupil-plane rms path-length error, the peak image concentration is then at $\lambda = 2\pi\sigma$. For ExAO-like correction levels of about 100 nm, the "ideal" observational wavelength from the direct imaging point of view is then in the red. In practice, note that both of these uses of a WCS are equally viable – in the first case, angular resolution is traded for a higher Strehl ratio at a given $\lambda$, while in the second case angular resolution is maintained by reducing $\lambda$.

---

[1]This same result also follows from the Fourier transform relationship between the pupil and focal plane fields that applies at large focal ratios (in this limit, the central value of the focal plane electric field is proportional to the average pupil-plane field, which, normalized, is 1-f, while the central intensity, proportional to the square of the field, is 1-2f, again reflecting the reductions due both to blockage and scattering), and also from an analytic treatment for the case of a central circular blockage (Born & Wolf 1975; Schroeder 1987).

While a significant increase in Strehl ratio is feasible with a WCS at any large telescope (e.g., Fig. 2), a readily accessible telescope is desirable for initial trials. In addition, since a telescope of diameter $\approx$ 5 m can reach the desired S > 90% regime (Fig. 2) where significant coronagraphic gains are possible, we decided to first implement a WCS on Palomar's 200 inch Hale telescope, for which the secondary blockage allows a maximum off-axis subaperture diameter of $\approx$1.5 m. This aperture diameter is comparable to aperture sizes proposed for several potential coronagraphic space telescopes (Trauger et al. 2005, Guyon et al. 2006), and is also 5/8 the size of the Hubble Space Telescope (HST).

Initial observations with the 1.5 m WCS provided by the sub-aperture relay optics bench (ROB) which we installed on the Hale telescope (Haguenauer et al. 2005, 2006; Serabyn et al. 2006), have now regularly provided excellent image quality. Fig. 4 shows an image of the single star HD121107 through our WCS. The observed stellar diffraction pattern closely resembles an ideal Airy pattern, with as many as 7 circular Airy rings quite prominent, and bits of rings up to the 9[th] evident. Longer integrations may have revealed even further Airy rings. The image is of course also free of linear diffraction structures because of the lack of pupil blockage. The best Strehl ratio obtained was $\approx$ 92-94%, corresponding to wavefront errors of 85 – 100 nm. Given experimental limitations in general (e.g., finite bandwidth and detector pixel size), the exact Strehl ratio is often somewhat uncertain (Roberts et al. 2004), but in our case the image quality is high enough that the main limitation to a precise Strehl ratio determination is likely its very small departure from unity. Such performance levels were not previously anticipated on large ground-based telescopes prior to the deployment of next-generation ExAO systems, although first generation AO systems on mid-sized telescopes have reached comparable levels in a few cases (Shelton et al. 1997, Oppenheimer et al. 2004). Indeed, the lack of diffraction spikes renders the image quality of Fig. 4 slightly higher than that of the HST's NICMOS instrument at the same wavelength (S = 90-92%; NICMOS handbook: www.stsci.edu/hst/nicmos), even without considering the additional 1-f scattering factor which applies to the HST case.

This high degree of wavefront correction has now also been verified at optical wavelengths. The B band (≈ 425 nm) was used for our initial visible observations, because the WFS removes essentially all of the longer-wavelength visible light. As our B-band image of the 0.34" separation binary SAO 37735 shows (Fig. 5), good AO correction is indeed present to wavelengths as short as the B-band, with both stellar images showing compact cores with widths of about 106 mas, or about 1.8 times the diffraction beam width. On an isolated star, SAO 85590, a B-band Strehl ratio of ≈ 0.12 was measured, again in accord with an rms of ≈ 100 nm (Fig. 3).

Such a WCS thus immediately enables a variety of experiments and observational modes previously thought to require next generation AO systems, including both infrared ExAO and visible AO, as we have demonstrated here, and high-contrast coronagraphy, which will be addressed in a subsequent paper. Potentially even predictive AO experiments can be carried out using additional wavefront information from beyond the selected sub-aperture. Of course even basic image differencing techniques (different wavelengths, polarizations and stars) should yield significantly better results with high and stable Strehl ratios. Fig. 6 shows the high Strehl ratio stability which our WCS can provide – in this data set spanning 140 sec, the mean Strehl ratio is 0.86, and its standard deviation is only 0.01. One measurement that can take advantage of such high PSF stability is the detection of the slight intensity changes caused by the transits and eclipses of hot planetary companions (Saumon et al. 1996). Such observations have recently been carried out from space (Charbonneau et al. 2005; Deming et al. 2005), but the very stable Strehl ratios evident in Fig. 6 should allow ground-based measurements of these effects as well.

Another potential experiment enabled by a WCS is the demonstration of a well-corrected "dark hole" near the central stellar image (Malbet, Yu and Shao 1995), useful for faint companion detection. Normally the diffracted and scattered starlight is too bright for a meaningfully dark hole to be generated with current generation AO systems. Fourier theory implies that the scattered light level in the domain of influence of an $N \times N$ element DM is roughly $(1-S)/N^2$, because the fraction of the light which is not in the well-corrected PSF core, 1-S, appears in a halo of size $N \times N$ spatial pixels. In our case of a $16 \times 16$ DM, $S > 0.9$ implies a scattered light level of $\approx 4 \times 10^{-4}$ relative to the star. This is deep enough for interesting brown dwarf searches very close to nearby stars, if the inner diffraction rings were also suppressed to this level.

As we point out here, the diffraction rings can theoretically be suppressed below this scattered light level simply by making use of the optical aberration known as coma. Because coma largely diffracts light to one side of the geometric image point, the opposite side of the PSF darkens (Born & Wolf 1984). For example, Fig. 7a shows the PSF resulting from the addition of 1.4 waves of coma to an otherwise perfect wavefront. In the cross-cut of Fig. 7b, it can be seen that all of the diffraction sidelobes on one side of the image are below $\approx 4 \times 10^{-4}$ of the aberrated peak, comparable to the predicted scattered light level. Note that such a dark-hole generation experiment requires no additional hardware, as it relies simply on adding a phase aberration term to the DM. The use of a more general pupil-plane phase distribution to generate even darker regions in the focal plane has been discussed by Codona et al. (2006).

Using our WCS, we recently carried out an initial attempt at such a coma experiment. We first took images of a single star while adding increasing amounts of coma to the DM, until a setting was found for which one side of the image was fairly dark. Next we observed the binary SAO 37434 with and without coma added. No optimization was attempted in this initial trial. Fig. 8 shows the resultant images. In the comatic image, the faint companion (down by a factor of $\approx$ 35), located just below center near the radius of the ideal $2^{nd}$ Airy ring, is much easier to distinguish than in the coma-free image, as would even fainter companions be. In the comatic image, the dark region near the companion is roughly $2 - 3 \times 10^{-3}$ as bright as the stellar peak. This rejection level is not as deep as predicted above, but is still well below the level of the ideal unaberrated $1^{st}$ Airy ring. This residual light level is presumably due to uncorrected low-order aberrations, as well as wavefront sensor errors aliased from higher spatial frequencies. The combination of a more optimized pupil phase map, the nulling of long-lived focal plane speckles (Bordé et al. 2006), and the use of a spatially filtered wavefront sensor (Poyneer and MacIntosh 2004) should lead to significantly improved performance.

A WCS thus provides a facility for immediately initiating a variety of ExAO experiments, as well as a practical near-term route for optimizing ExAO and high-contrast systems well in advance of the advent of costly ExAO systems on large telescopes. Indeed, a potential route to accelerating the development of two-stage ExAO systems on large telescopes is to match a sub-aperture to a smaller and more readily available DM in the second stage, until a full-scale DM is available. However, while a WCS has great potential as a facility for carrying out ExAO demonstration experiments, it is also of interest in its own regard. Most obviously, a WCS immediately enables visible wavelength AO observations. A WCS could also immediately be implemented on existing 8-10 m class telescopes to provide WCSs as large as ≈ 4 m. This is a size potentially large enough to allow the detection of bright, self-luminous extra-solar jovian planets (Saumon et al. 1996), if the stellar wavefront were corrected and rejected well enough. Beyond the current era, proposed telescopes of diameter 30 – 40 m could potentially provide clear WCSs of order 15 m, large enough to potentially detect favorable reflected light planets around the nearest stars (Angel 1994), and a large off-axis telescope (e.g. Storey et al. 2006) is also an attractive possibility in this regard. The main question is then the relative coronagraphic performance that can be achieved with a clear off-axis WCS, as opposed to with the more poorly corrected and partially vignetted full telescope pupil. The small inner working angles that some recently proposed coronagraphs theoretically possess (Guyon 2006) may be well-suited to making use of a smaller, but better corrected telescope aperture, and as mentioned earlier, a FQPM coronagraph actually requires an unvignetted circular aperture for high rejection. Indeed, the exchange of aperture area for higher measurement accuracy is already familiar from the case of aperture masking (e.g., Lloyd et al. 2006). The potential coronagrapic performance of a WCS will thus be the topic of our next paper.

**Acknowledgements**

This work was carried out at the Jet Propulsion Laboratory, California Institute of Technology, under contract with the National Aeronautics and Space Administration, and is based in part on observations obtained at the Hale Telescope, Palomar Observatory, as part of a continuing collaboration between the California Institute of Technology, NASA/JPL, and Cornell University. We thank the staff of the Palomar Observatory for their able and ready assistance, and the JPL Research and Technology Development program for funding this work. Finally, we thank the referee, who's name cannot be spoken, for numerous helpful comments.

**Figure Captions**

Figure 1: Conceptual relay optics layout. The pair of paraboloids f1 and f2 magnify the pupil by the ratio of their focal lengths, f1/f2, while the field lens F maintains the pupil location on the DM. A stop at pupil location P1 transmits a linear fraction f2/f1 of the pupil.

Figure 2: a) Expected Strehl ratios at 2.2 microns for the full aperture (lower curve) and 1.5 m clear off-axis subaperture (upper curve) cases on the Hale telescope. A 16 x 16 element Shack-Hartman WFS is assumed in both cases. b) Expected Strehl ratios at 2.2 microns for the 10 m aperture (lower curve) and 4 m clear off-axis subaperture (upper curve) cases for a 10 m telescope on Mauna Kea. A 20 x 20 element Shack-Hartman WFS is assumed in both cases.

Figure 3: Expected Strehl ratio vs. wavelength for 250, 200, 100 and 80 nm of rms wavefront error (bottom to top). The two lower curves give a range approximating that of current AO performance levels, and the upper two curves roughly delineate a potential ExAO regime attainable with next generation systems. The vertical line is at the wavelength of H$\alpha$, and the horizontal line is at a Strehl ratio of 0.9, beyond which coronagraphic gains become significant.

Figure 4: Image of the single star HD121107 obtained with our 1.5 m off-axis WCS on the Hale telescope on the night of 14 June 2005. The passband is defined by the 1% wide "Br $\gamma$" filter centered at 2.17 µm. This image is the sum of 20 short-exposure individual images, with a total integration time of 28 s.

Figure 5: B-band image of the binary SAO 37735 obtained with our WCS on the night of 10 Sep. 2006. The pixel size is 25.3 mas, $\lambda/D$ is $\approx$ 59 mas, and the binary separation is 0.34". The V magnitudes of the two stars are 5.1 and 6.3.

Figure 6. Sequence of Strehl ratio measurements for one hundred consecutive 1.4 sec exposures obtained on the night of 13 June 2005 on $\alpha$ Boo.

Figure 7. Top: Calculated PSF for the case of 1.4 waves of coma (peak to peak in a 1.5 m aperture) added to an otherwise perfect plane wave. Bottom: A horizontal cross-cut through the comatic PSF. All of the sidelobes on the right-hand side of the image are $< 4 \times 10^{-4}$ of the aberrated peak (which is about 0.7 of the unaberrated peak).

Figure 8. Left: Image of the binary SAO37434 (separation = 0.78", ratio = 35:1) in the $K_s$ band, obtained through our Palomar WCS on the night of 9 Sep. 2006. Right: Image of the same binary through our WCS with about 1 wave of vertical coma added to the DM to darken the lower half plane. The fainter of the pair of stars is marked by an arrow.

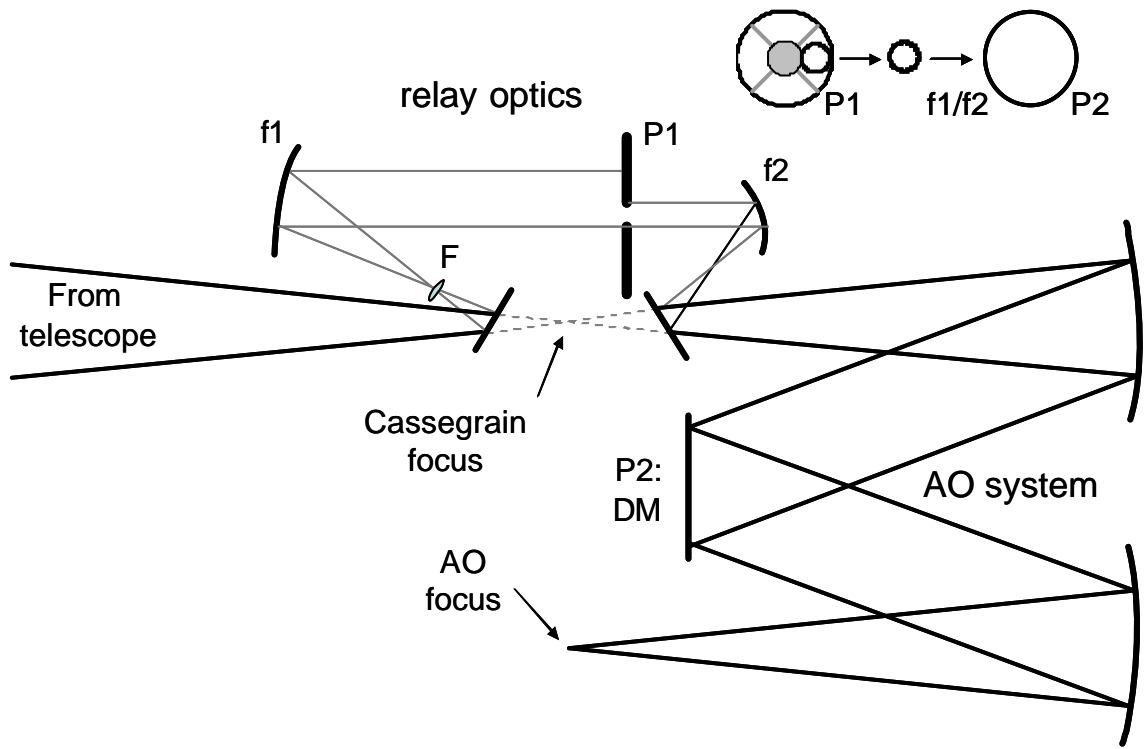

Figure 1

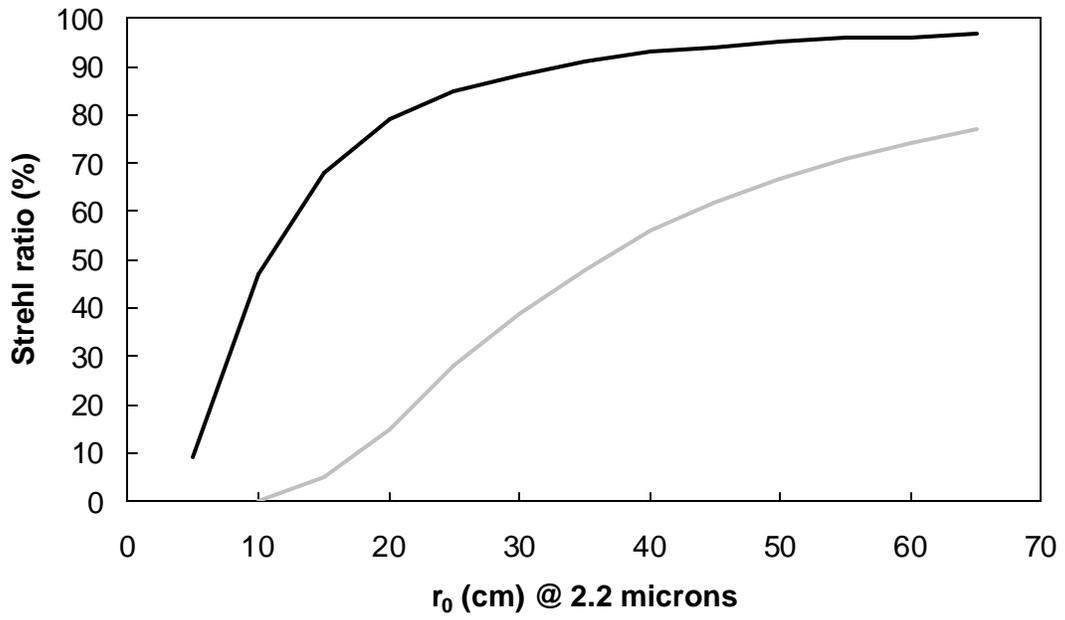

Figure 2a

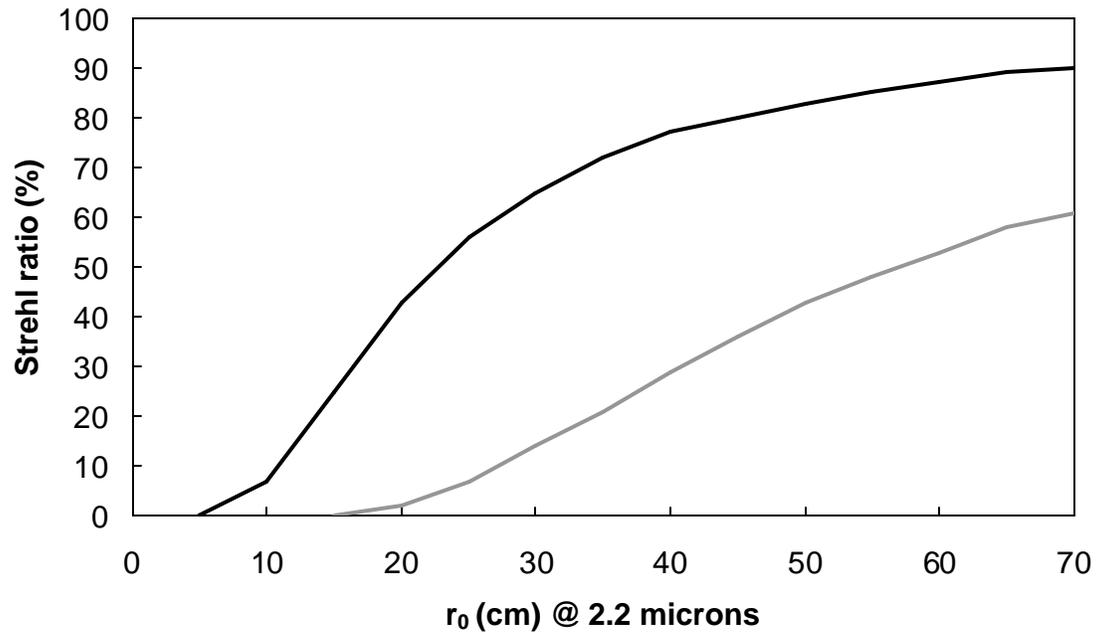

Figure 2b

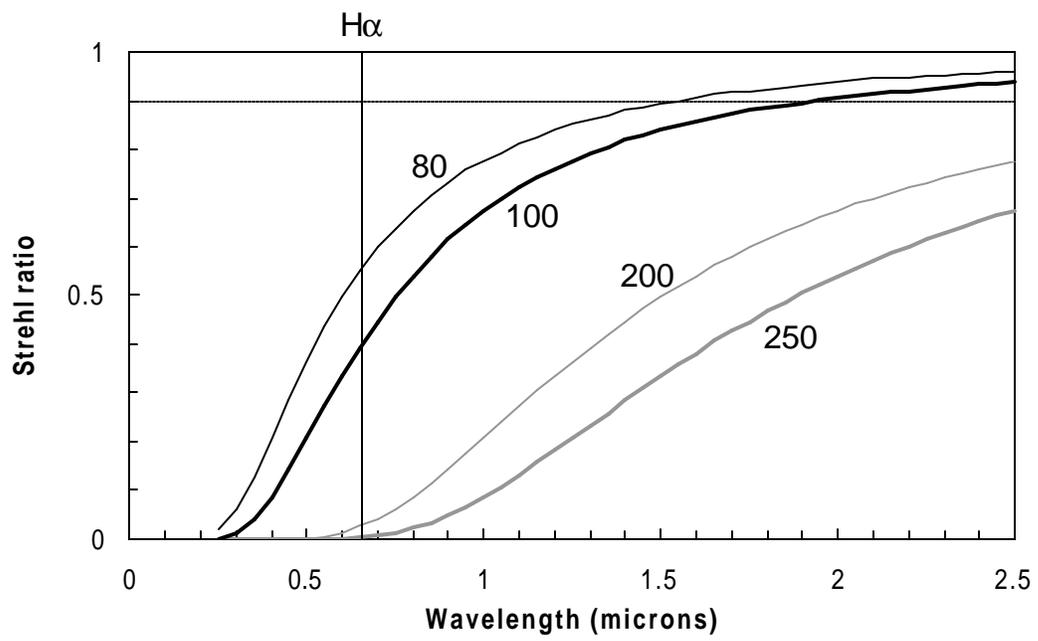

Figure 3

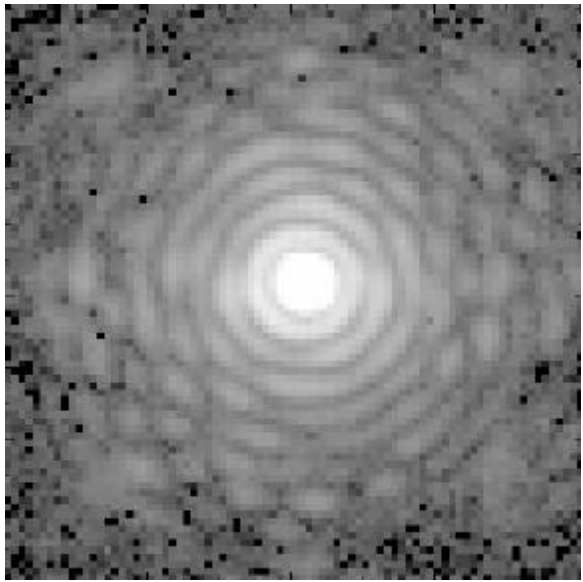

Figure 4

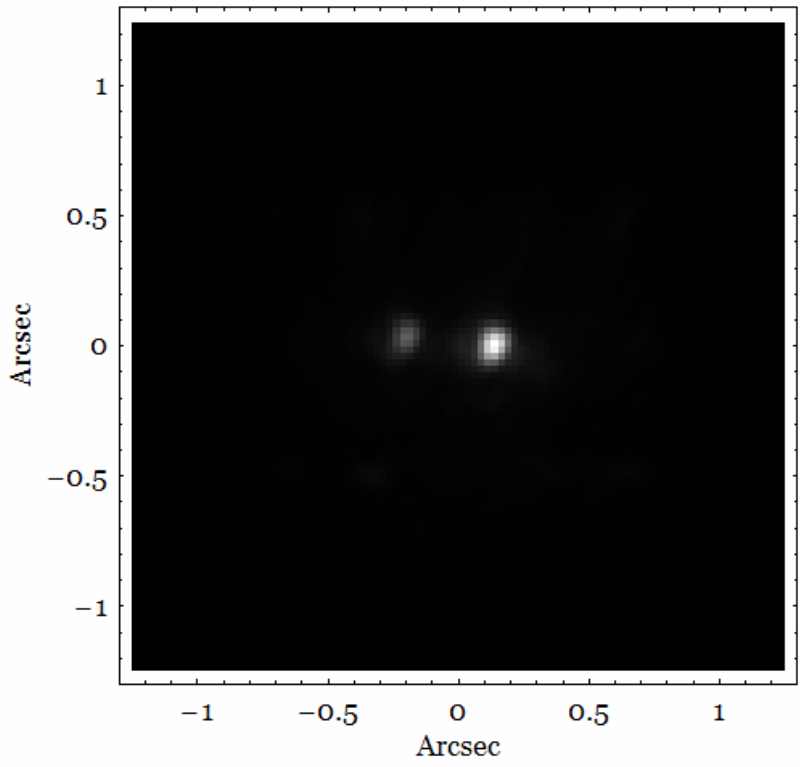

Figure 5

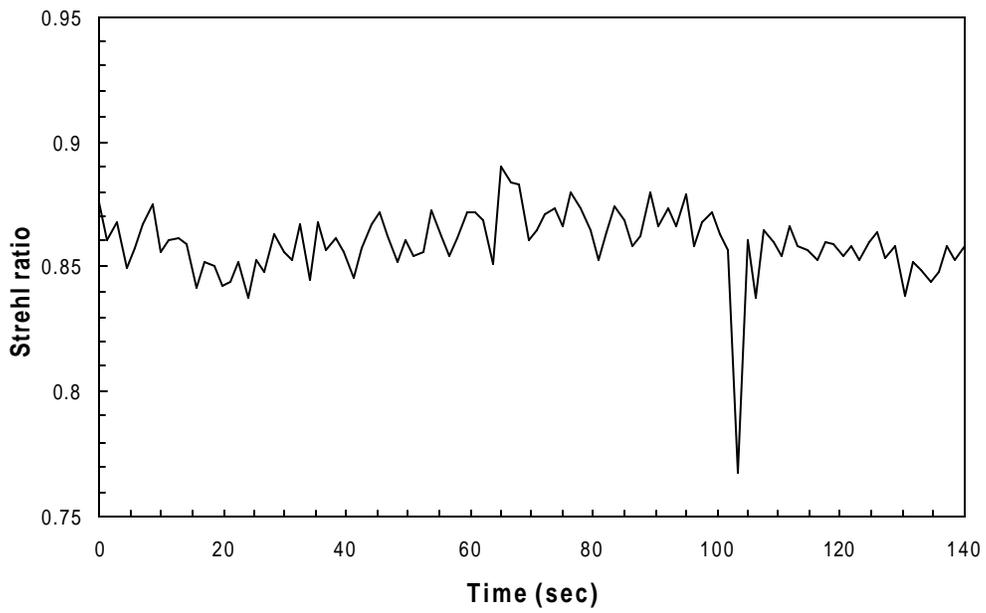

Figure 6

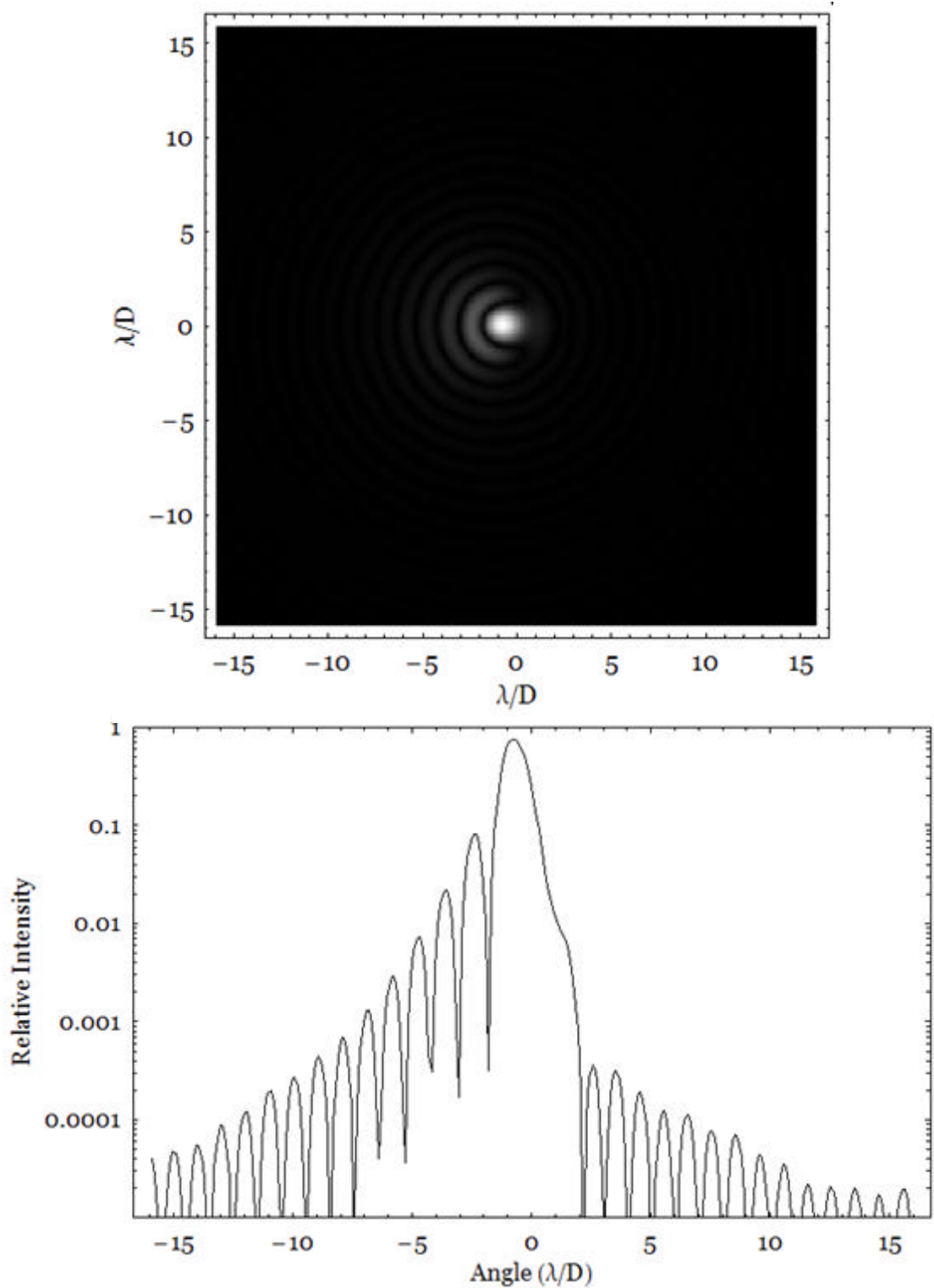

Figure 7

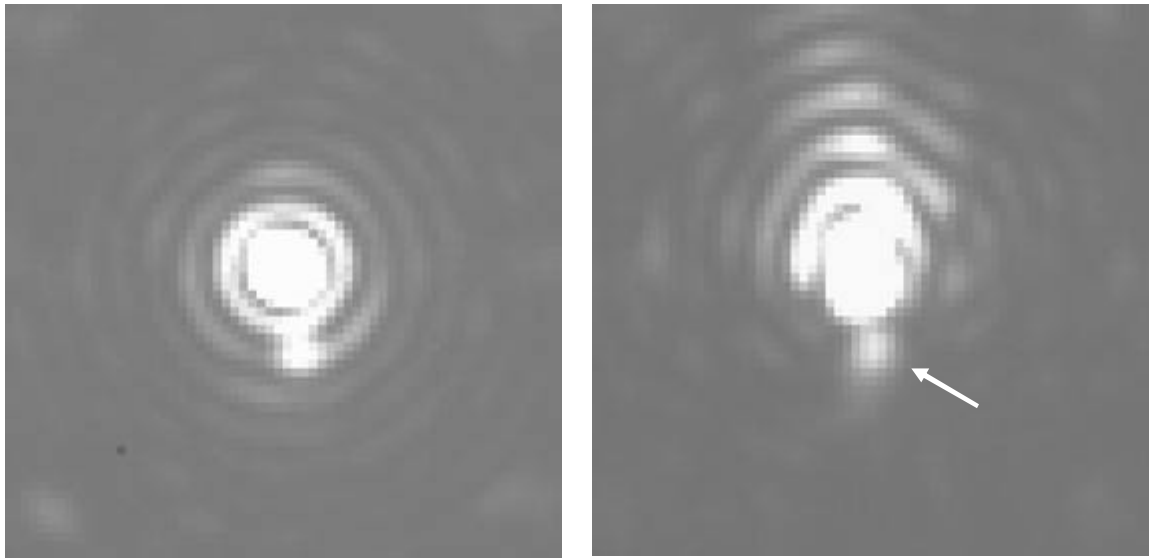

Figure 8